\documentclass[preprintnumbers,floats,twocolumn,prd,aps,superscriptaddress]{revtex4-1}
\usepackage{latexsym}
\usepackage{amsfonts}
\usepackage{psfrag}
\usepackage{graphicx}
\usepackage{bm}
\usepackage{amssymb,amsmath,epsfig}
\usepackage{subfig}
\usepackage{float}
\usepackage[titletoc,title]{appendix}

\begin{document}

\setlength{\pdfpagewidth}{8.5in}

\setlength{\pdfpageheight}{11in}

\title{Reinventing the Zel'Dovich Wheel}
% The Zel'Dovich Effect Revisited
% A Reformulation of the Zel'Dovich Effect

\author{Cisco Gooding}%\email{cgooding@physics.ubc.ca}
\affiliation{
	School of Mathematical Sciences, 
	University of Nottingham, UK
	}
\affiliation{
  Department of Physics \& Astronomy,
  University of British Columbia, Canada
}
\author{Silke Weinfurtner}%\email{Silke.Weinfurtner@nottingham.ac.uk}
\affiliation{
	School of Mathematical Sciences, 
	University of Nottingham, UK
}
\affiliation{
	Centre for the Mathematics and Theoretical Physics of Quantum Non-Equilibrium Systems, 
	University of Nottingham, UK
}
\author{William G. Unruh}%\email{unruh@physics.ubc.ca}
\affiliation{
  Department of Physics \& Astronomy,
  University of British Columbia, Canada
}
\date{\today}

\begin{abstract}
After reviewing the pioneering work by Zel'Dovich in which radiation is amplified perpendicular to the axis of a rotating conductor, we consider an alternative scattering arrangement. We demonstrate superradiant amplification of electromagnetic waves with orbital angular momentum, directed axially towards a rotating conductor. Taking into account recent advances in optics and condensed matter systems, our approach presents new possibilities for rotational superradiance experiments. We discuss remaining challenges that must be faced before a laboratory observation of the Zel'Dovich effect can be considered feasible.
\end{abstract}

\maketitle

\section{Introduction}

Rotational superradiance was first demonstrated theoretically in the seminal work of Zel'Dovich \cite{Zeldovich71,Zeldovich72}, where it was found that rotating, conducting cylinders can superradiate. A crucial requirement of the conducting cylinder is that it not be a perfect conductor, but have a finite conductivity to absorb radiation entering the cylinder. Zel'Dovich's formulation involved sending an electromagnetic wave perpendicular to the rotation axis of the spinning conductor, and subsequent work on superradiance has utilized the same configuration. 

Since conception, observing the Zel'Dovich effect in the laboratory has been considered an unlikely prospect at best. The difficulty stems largely from the superradiance condition (given explicitly in the following section), which requires either the use of very rapid rotation rates or very long wavelengths. Due to the small size of the wave amplification \cite{Zeldovich71,Zeldovich72}, and the practical difficulty with how Zel'Dovich's reflected waves scatter from the conducting cylinder in all radial directions, neither the rapid-rotation nor the long-wavelength regime render the amplification observable.

%%Rapid rotation rates constrain the size of the conductor: for a given rotation rate, increasing the conductor radius increases both the radial tensile load and the tangential velocity at the outer surface. Hence, the conductor will break apart if rotated rapidly enough, due to structural failures or shock wave instabilities.  long-wavelength regime is impractical 

To combat such difficulties, people have turned to analogue systems that can also exhibit superradiance, but have much lower wave speeds and more convenient dispersion relations. In particular, the first experimental observation of rotational superradiance was recently achieved using surface water waves on a vortex flow \cite{Silke2017}. 

Alternatively, in this work we consider a reformulation of the original Zel'Dovich effect. As we have previously shown \cite{GWU2018}, there exists a previously unexplored direction for superradiant amplification that offers promising simplifications to the usual scattering arrangement. Our approach involves sending the wave parallel to the conducting cylinder's rotation axis and reflecting off the flat ends of the cylinder. This new alignment reduces the effective dimensionality of the scattering system, simplifying both the theoretical description and the corresponding experimental setup.

Using the formalism of relativistic electrodynamics, we demonstrate that classical amplification, and therefore stimulated emission, can indeed occur in our proposed scattering configuration. The electromagnetic modes shown to be amplified carry orbital angular momentum (OAM), and are directed along the rotation axis. Since Einstein showed over a century ago that stimulated emission implies spontaneous emission \cite{EinsteinAB}, our findings also imply that rotating conductors will spontaneously emit OAM-carrying photons along the rotation axis.

We then discuss how our proposal could be implemented experimentally by shining an OAM-carrying laser at a rapidly rotating conductor. As we describe in detail, such an implementation is difficult, due to the large rotation rate required to satisfy the superradiance condition. Nonetheless, recent advancements in optics and condensed matter have led to unprecedented rotation rates. At the microscale, dumbbell rotators have achieved $\text{GHz}$ rotations \cite{GHzRotor,GHzRotor2}. Such high rotation rates indicate that these microscale rotators might be used to observe superradiance. 

Even at microscales, the rotators are large enough to behave essentially as classical objects. Still, in analogy with superradiance of modes from a test field scattered from a rotating black hole \cite{Unruh2ndQuantization}, such rotators could potentially allow quantum aspects of rotational superradiance to be observed for the first time, provided coherence can be maintained in the laser field. Our proposal therefore offers a complementary approach to the study of both classical and quantum superradiance. 

\section{Relativistic Electrodynamics}

We work within the formalism of relativistic electrodynamics in curvilinear coordinates, following the approach of Bekenstein and Schiffer \cite{Faces}. The rotating conductor is modelled as a conducting dielectric with spatially uniform permittivity $\epsilon(\omega)$ and permeability $\mu(\omega)$, which we will take to both be real.

The main quantity of interest will be the electromagnetic field tensor $F^{\mu\nu}$, which is composed in an observer-dependent way of the electric field $\bm{E}$ and magnetic induction $\bm{B}$. We will also be interested in the electromagnetic displacement tensor $H^{\mu\nu}$, which is composed of the electric displacement $\bm{D}=\epsilon\bm{E}$ and the magnetic field $\bm{H}=\mu^{-1}\bm{B}$. Along with Ohm's law $\bm{j}=\sigma\bm{E}$ (with $\sigma$ being the conductance), these relations can be written in tensorial form as
\begin{eqnarray}\label{Const1}
H^{\alpha\beta}u_\beta=\epsilon F^{\alpha\beta}u_\beta \\
*\, F^{\alpha\beta}u_\beta=\mu * H^{\alpha\beta}u_\beta \\
j^\alpha = \sigma F^{\alpha\beta}u_\beta + \rho u^\alpha,\label{Const3}
\end{eqnarray}
where $\rho$ is the proper charge density and $u^\alpha$ is the $4$-velocity of the material. The $*$ symbol is the Hodge star operator, defined such that $*\, F^{\alpha\beta}\equiv \frac{1}{2}\varepsilon^{\alpha\beta\gamma\delta}F_{\gamma\delta}$, with $\varepsilon^{\alpha\beta\gamma\delta}$ denoting components of the Levi-Civita tensor (sign convention: $\varepsilon^{0123}=+1$). The electric and magnetic fields are defined with respect to the rest frame of the object, and the frequency-dependent quantities $\epsilon(\omega)$ and $\mu(\omega)$ are defined with respect to this rest frame as well.

We use a cylindrical coordinate system $\{x^\mu\}=\{t,r,\phi,z\}$, in which case the Minkowski metric takes the form
\begin{equation}
ds^2=-dt^2+dr^2+r^2 d\phi^2+dz^2.
\end{equation}
Inside the conductor, the co-rotating $4$-velocity has components
\begin{equation}
u_\alpha = \gamma(-1,0,\Omega r^2,0),
\end{equation}
with the Lorentz factor $\gamma=1/\sqrt{1-\Omega^2 r^2}$. The relations (\ref{Const1})-(\ref{Const3}) imply
\begin{eqnarray}
\epsilon^{-1}H^{02}=F^{02}\equiv r^{-1}E_\phi \\
\mu H^{31}=F^{31}\equiv B_\phi \\
\epsilon^{-1}\left(H^{01}+\Omega r^2 H^{12}\right)=F^{01}+\Omega r^2 F^{12}\equiv \gamma^{-1}E_r \\
\mu\left(H^{23}-\Omega H^{03}\right)=F^{23}-\Omega F^{03}\equiv (r\gamma)^{-1}B_r \\
\epsilon^{-1}\left(H^{03}-\Omega r^2 H^{23}\right)=F^{03}-\Omega r^2 F^{23}\equiv \gamma^{-1}E_z \label{LattConst1}\\
\mu\left(H^{12}+\Omega H^{01}\right)=F^{12}+\Omega F^{01}\equiv \left(r\gamma\right)^{-1}B_z.\label{LattConst2}
\end{eqnarray}

Maxwell's equations are given by
\begin{eqnarray}\label{MaxHom}
F_{[\alpha\beta,\gamma]}=0, \\
\left(H^{\alpha\beta}\right)_{,\beta}=4\pi j^\alpha.\label{MaxInhom}
\end{eqnarray}
For electromagnetic field modes with temporal dependence $e^{-i\omega t}$ and azimuthal dependence $e^{im\phi}$ with respect to the lab frame, the homogeneous equation (\ref{MaxHom}) implies
\begin{eqnarray}
i\omega r^2 F^{12}-\partial_r \left(F^{02}r^2\right)+im F^{01}=0 \\
\partial_r \left(F^{23}r^2\right)+im F^{31}+r^2 \partial_z F^{12}=0 \\
i\omega F^{31}+\partial_r F^{03}-\partial_z F^{01}=0 \\
i\omega r^2 F^{23}-im F^{03}+r^2\partial_z F^{02}=0,
\end{eqnarray}
while the inhomogeneous equation (\ref{MaxInhom}) implies
\begin{align}
&\partial_r\left(H^{01}r\right)+i m r H^{02}+r\partial_z H^{03}=4\pi r\gamma\left(\sigma\Omega r E_\phi+\rho\right) \\
&i\omega H^{01}+im H^{12}+\partial_z H^{13}=4\pi \sigma E_r \label{IH2}\\
&i\omega r H^{02}-\partial_r \left(H^{12}r\right)+r\partial_z H^{23}=4\pi\gamma\left(\sigma E_\phi+r\Omega\rho\right) \\
&i\omega r H^{03}+\partial_r \left(H^{31}r\right)-i m r H^{23}=4\pi\sigma r E_z.\label{IH4}
\end{align}
These equations are corrected from the forms given by Bekenstein and Schiffer in \cite{Faces}; the differences include a minus sign on the third term on the left-hand side of (\ref{IH2}), and a factor of $r$ on the right-hand side of (\ref{IH4}). Note also that we are not scaling the charge density by $4\pi$, as was done in \cite{Faces}.

We will approximate the electromagnetic field modes as transverse, such that $B_z=E_z=0$. In practice, laser modes with orbital angular momentum will have longitudinal components, but these components are small, and will be neglected in this work. The constitutive relations (\ref{LattConst1})-(\ref{LattConst2}) then imply $F^{23}=\mu H^{23}=\gamma B_r/r$, $\mu H^{03}=F^{03}=r\Omega\gamma B_r$, $\epsilon^{-1}H^{01}=F^{01}=\gamma E_r$, and $F^{12}=\epsilon^{-1}H^{12}=-\Omega\gamma E_r$. Using these relations, the homogeneous Maxwell's equations allow for the determination of $F^{01}$ and $F^{03}$, in terms of $F^{02}$:
\begin{equation}
F^{01} = \frac{i \partial_r \left(F^{02}r^2\right)}{\left(\omega \Omega r^2-m\right)},
\end{equation} 
\begin{equation}
F^{03} = \frac{i\Omega r^2\partial_z F^{02}}{\left(\omega-m\Omega\right)}.
\end{equation}
From $F^{01}$ and $F^{03}$, $E_r$ and $B_r$ are determined, respectively. The homogeneous Maxwell's equations also allow $F^{31}$ (and therefore $B_\phi$) to be solved for in terms of $F^{02}$:
\begin{equation}
F^{31} = \left[\gamma^2 \left(\omega \Omega r^2-m\right)\left(\omega-m\Omega\right)\right]^{-1}\partial_z \partial_r \left(F^{02}r^2\right).
\end{equation}

The only remaining degree of freedom is $F^{02}$ (equivalently, $E_\phi$), upon which all other field components depend. Combining the inhomogeneous Maxwell's equations judiciously, we obtain
\begin{equation}\label{F02}
\partial_z^2 F^{02}+\left(V+i\Gamma\right)F^{02}=0,
\end{equation}
with the definitions 
\begin{equation}
V = \mu \epsilon \gamma^2 \left(\omega-m\Omega\right)^2 \hspace{6pt} \text{and} \hspace{6pt}
\Gamma = \mu \gamma 4\pi \sigma \left(\omega-m\Omega\right).
\end{equation}
One can also show that $F^{01}$ obeys the same isolated dynamical equation as $F^{02}$ does, in a manner analogous to the derivation of (\ref{F02}):
\begin{equation}\label{F01}
\partial_z^2 F^{01}+\left(V+i\Gamma\right)F^{01}=0,
\end{equation}
with the same $V$ and $\Gamma$ as in (\ref{F02}).

Next we demonstrate that the system can superradiate. We will show that if the transverse electromagnetic field is incident on the top ($z=z_0$) of the rotating conductor, there will be a net positive longitudinal energy flux coming from the top of the conductor, provided the superradiance condition $\omega-m\Omega<0$ is satisfied. Energy flow is characterized by the Poynting vector, $\bm{S}=(\bm{E}\times\bm{H})/4\pi$; we will specifically be interested in the $z$ component, which we can express as
\begin{equation}\label{EMPoynting}
S_z=\frac{1}{4\pi}\left(\bm{E}\times\bm{H}\right)_z=\frac{1}{4\pi}\left(F^{01}H^{31}-F^{02}H^{23}\right).
\end{equation}

In terms of complex field amplitudes, the time-average of the Poynting vector is $\bar{\bm{S}}=\text{Re}\left[(\bm{E}\times\bm{H}^*)_z/8\pi\right]$. Using the constitutive relations and the homogeneous Maxwell equations, we can then write the axially-directed energy flow (\ref{EMPoynting}) within the conductor in time-averaged form as
\begin{align}
\bar{S}_z=&\frac{1}{8\pi\mu\left(\omega-m\Omega\right)}\\
&\cdot \text{Re}\left[\frac{1}{\gamma^2}iF^{01}\partial_z\left(F^{01}\right)^*+iF^{02}\partial_z\left(F^{02}\right)^*\right].\nonumber
\end{align}

Outside of the conductor, in the $z>z_0$ region, we have
\begin{equation}\label{ExtFlux}
\bar{S}_z(z>z_0)=\frac{1}{16\pi\omega}\left[iW_1(z)+iW_2(z)\right],
\end{equation}
with the definition
\begin{equation}\label{jWronskian}
W_j(z)=F^{0j}\partial_z\left(F^{0j}\right)^*-\left(F^{0j}\right)^*\partial_z F^{0j},
\end{equation}
for $j\in\{1,2\}$. The expression (\ref{jWronskian}) gives the Wronskians of equations (\ref{F02}) and (\ref{F01}) outside of the cylinder. In this region the Wronskians have vanishing $z$-derivatives, so we can find the longitudinal energy flow far from the conductor by evaluating $W_1$ and $W_2$ in (\ref{ExtFlux}) at $z_0$ (approaching $z_0$ from above).

Inside the conductor, the quantities defined by (\ref{jWronskian}) are still Wronskian-like \cite{GenSuperradiance}, and satisfy
\begin{equation}\label{InWronskian}
\frac{\partial}{\partial z}\left( iW_j(z)\right)=-2\Gamma |F^{0j}|^2.
\end{equation}
We assume that the field does not penetrate the back of the disk; hence, there will be a negligible field in the $z<0$ region. Upon integration of (\ref{InWronskian}) from $z=0$ to $z=z_0$, we then find
\begin{equation}
iW_j(z_0)=-2\int_0^{z_0}dz\,\Gamma |F^{0j}|^2.
\end{equation}
By exploiting the continuity of the Poynting vector at $z=z_0$, we now have
\begin{equation}\label{BarS}
\bar{S}_z(z>z_0)=\frac{-\sigma\mu\left(\omega-m\Omega\right)r\gamma}{2\omega}\int_0^{z_0}dz\,\left(|F^{01}|^2+|F^{02}|^2\right).
\end{equation} 
The entire energy flux out of the cylinder in the $z$ direction can be obtained by integrating (\ref{BarS}) over the circular cross section. Since the second law of thermodynamics implies that the conductivity is greater than or equal to zero \cite{Faces}, we can conclude from expression (\ref{BarS}) that there will be a positive net longitudinal energy flux far from the conductor whenever $\omega-m\Omega<0$; in other words, the system exhibits superradiance.

\section{Discussion}

We have demonstrated that an OAM-carrying electromagnetic wave directed axially at a rapidly rotating conductor can superradiate. This offers an alternative scattering arrangement to observe rotational superradiance in an electromagnetic system, which complements the original formulation first suggested by Zel'Dovich \cite{Zeldovich71,Zeldovich72}. 

The scattering arrangement considered here provides a number of benefits, compared to the usual scheme. In particular, due to the axial nature of our scattering setup, the dimensionality of the system is effectively reduced by one: directed beams reflect off the face of a rotating conductor back towards their source, instead of scattering off the sides in all directions (perpendicular to the rotation axis). This makes the experimental setup simpler, and ideal for implementation with lasers/masers.

Several technical challenges must be faced before the theoretical effect demonstrated here could be observed experimentally. As mentioned in the Introduction, the main constraint comes from satisfying the superradiance condition, $\omega-m\Omega<0$. The beam frequency is restricted by the size of the associated wavelength, since this also sets the scale for the spatial extent of the conductor; consequently, the minimum beam frequency for a reasonable experimental setup would be roughly a $\text{GHz}$. Generating an OAM beam in that frequency range can be accomplished in a variety of ways \cite{MilliLG1996,GyrotronMasers2017,LowFreqRadio2007,OAMRadio,GHzRadio,THz}. It then follows that the product of $m$ and $\Omega/2\pi$ would need to be of the order $10^{9} \text{Hz}$ (or larger) to be in the superradiant regime. 

We would therefore need to rotate the conductor as fast as possible. Observations of frequency shifts in scattered laser light due to a rotating body, referred to as the ``rotational Doppler effect,'' have typically involved objects much larger than microscale and rotation rates less than a $\text{kHz}$ \cite{Doppler,AngularAccel,RotDopplerNonlinearOpt,RotFreqShiftNew,RotFreqShiftmm,RotFreqShiftMeas}. One could in principle attain significantly higher rotation rates. However, for larger than microscale conductors, the rotation rate could not exceed $10^5\text{Hz}$; such rapid rotation rates have been achieved by dentist drills, for instance, which small conductors can be mounted on. This would in turn require a topological charge for the incident beam of $m=10^4$. Topological charges of $10^4$ have been generated \cite{Anton10000}, so our proposal could be experimentally feasible at this scale, provided there is sufficient coupling between the OAM beam and the conductor \cite{Couple}.

Another difficulty that arises comes from how the radial intensity profile of typical OAM beams changes with increasing OAM. One usually finds that the intensity of an OAM beam is negligible along the propagation axis, and is localized for the most part at some finite radius. The superradiance demonstration presented above assumes the disk has a radius at least as large as the radial extent of the incident OAM beam; however, since increasing the OAM of a beam typically increases the radius of maximum intensity \cite{SmallBeamHighTopo}, for a rotator with finite disk radius, increasing OAM can lead to an appreciable reduction of overlap between the incident beam and the rotator. In fact, an argument given by Wald and Mackewicz (private communication) indicates that this overlap will be small whenever the following are all met: $(i)$ the mode is propagating, $(ii)$ the mode satisfies the paraxial condition, $(iii)$ the tangential velocity of the disk at $r=R$ is less than the speed of light, and $(iv)$ the superradiance condition is satisfied. To prove this, we note that the paraxial approximation requires the beam to be highly collimated, such that $(m/\omega^2 w_0^2)\ll 1$. Now, the maximum intensity radius for a propagating Laguerre-Gaussian paraxial mode in the focal plane is $\sqrt{m/2}w_0$, where $w_0$ is the beam waist for the $m=0$ mode. Hence, for maximal overlap with the disk, one must have $\sqrt{m/2}w_0<R$. For the tangential velocity at $r=R$ to be less than the speed of light (recall $c=1$, here), we impose $R\Omega<1$. Then, applying these constraints successively,
\begin{equation}\label{Wald}
1\gg \frac{m}{\omega^2 w_0^2}>\frac{m^2}{2\omega^2 R^2}>\frac{m^2 \Omega^2}{2\omega^2}>\frac{1}{2},
\end{equation}
where the superradiance condition $\omega-m\Omega<0$ was used to yield the rightmost inequality. 

One can conclude from (\ref{Wald}) that if conditions $(i)-(iv)$ are met, adding the additional requirement of $\sqrt{m/2}w_0<R$ leads to a contradiction. This argument is related to the argument given in \cite{GWU2018} for acoustic systems, which shows that one can either amplify propagating incident modes and have the outer edge of the rotating absorber travel faster than the sound speed, or amplify evanescent modes and keep the rotator motion below the sound speed. In the current electromagnetic context, we are left in a similar position: we can either amplify evanescent OAM modes (using a waveguide, for instance), or we can amplify non-paraxial propagating modes. For the latter possibility, 
large OAM beams would still have to be significantly focused to interact with a small spinning conductor.

Such strong focusing has a large effect on the structure of the electromagnetic field, producing in particular a non-negligible field in the direction of propagation \cite{SmallBeamHighTopo2}. Complications associated with the strong focusing could be avoided altogether by working with the non-paraxial modes known as perfect optical vortices: OAM beams with diameters that are independent of the OAM \cite{PerfectHighTopo,FirstPerfect,SimplePerfect,PerfectHighOrder,EffPerfVort2017,PerfectVectorScalarVortex}. The independence of the beam diameter with respect to OAM allows higher OAM modes to be used for the same size rotators, without changing the setup. However, it is not yet clear if these perfect optical vortices can be prepared with small enough beam diameters to be applied to nanoscale or even microscale rotators.

The microscale dumbbell rotators mentioned in the Introduction \cite{GHzRotor,GHzRotor2} operate in the $\text{GHz}$ rotation range, and could thus be used in conjunction with lasers/masers with significantly lower topological charges. This would alleviate the experimental burden of involving beams with such high OAM. Since the dumbbells are not rotated about their symmetry axis, our assumption of cylindrical symmetry is strictly speaking not satisfied. Similarly, our analysis also ignores edge effects that can arise due to the finite size of the rotators (this is especially relevant if the wavelength is sufficiently larger than the size of the object). Nonetheless, scattering cross-sections for asymmetric scattering centers are often estimated by calculations assuming azimuthal symmetry; likewise, edge effects are often neglected at the lowest order of approximation, and corrected for perturbatively. Hence, we expect such differences to affect only the characteristics of how they superradiate, and not prevent superradiance from occurring altogether.

Once it becomes possible to observe superradiant amplification of electromagnetic signals from a rotating conductor at the classical level, one can begin working towards observing quantum aspects of the effect. Current optomechanical experiments can demonstrate entanglement between the OAM in laser modes and the ground-state angular oscillations of torsion pendula \cite{RovibeEntangle2008,LaserRot2007,RotoMirrorEntangle2008,RotoStabilization,LargeAngularMom}. At the nanoscale, rotators have been prepared showing rotational coherence in the $\text{THz}$ rotation range \cite{THzRotor,MacroRotSup}. Preliminary steps have also been taken to entangle the OAM of photons carrying more than $10^4$ angular momentum quanta with the polarization of partner photons \cite{Anton10000}, paving the way for further developments in coupling individual high-OAM photons to mechanical oscillators.

Amplifying electromagnetic signals via rotational superradiance has long been dismissed as a practically unobservable theoretical curiosity. However, despite the experimental challenges involved in implementing our proposal, the approach described here is geometrically simpler than the original Zel'Dovich setup, and represents a new avenue for theoretical and experimental superradiance research.

\section{Acknowledgements}

The authors thank Bob Wald and Kris Mackewicz for pointing out the difficulty with amplifying propagating paraxial modes. The research of WGU is supported by NSERC (Natural Science and Engineering Research Council) of Canada, and also by CIfAR. SW acknowledges financial support provided under the Paper Enhancement Grant at the University of Nottingham, the Royal Society University Research Fellow (UF120112), the Nottingham Advanced Research Fellow (A2RHS2), the Royal Society Project (RG130377) grants, the Royal Society Enhancement Grant (RGF/EA/180286) and the EPSRC Project Grant (EP/P00637X/1). SW acknowledges partial support from STFC consolidated grant No. ST/P000703/. The fellowship held by CG while this research was conducted was funded by NSERC through WGU, with partial support from SW.


\begin{thebibliography}{99}

\bibitem{Zeldovich71}
  Y.~B.~Zel'Dovich,
	``Generation of Waves by a Rotating Body,''
	Zh.\ Eksp.\ Teor.\ Fiz.\ {\bf 14}, 270 (1971) [JETP Lett.\ {\bf 14}, 180 (1971)].

\bibitem{Zeldovich72}
  Y.~B.~Zel'Dovich,
  ``Amplification of Cylindrical Electromagnetic Waves Reflected from a Rotating Body,''
	Zh.\ Eksp.\ Teor.\ Fiz.\ {\bf 62}, 2076 (1972) [Sov.\ Phys.\ JETP {\bf 35}, 1085 (1972)].

\bibitem{Silke2017}
  T.~Torres, S.~Patrick, A.~Coutant, M.~Richartz, E.~W.~Tedford, and S.~Weinfurtner,
	``Rotational superradiant scattering in a vortex flow,''
	Nature Physics {\bf 13}, 833 (2017).

\bibitem{GWU2018}
  C.~Gooding, S.~Weinfurtner, and W.~G.~Unruh,
	``Superradiant scattering of orbital angular momentum beams,''
	(in review). Preprint: arXiv:1809.08235 (2018).
	
% Einstein A and B paper
\bibitem{EinsteinAB}
  A.~Einstein,
	``The Quantum Theory of Radiation,''
	Phys.\ Z.\ {\bf 18}, 121 (1917).		
	
\bibitem{GHzRotor}
  J.~Ahn, Z.~Xu, J.~Bang, Y.~-H.~Deng, T.~M.~Hoang, Q.~Han, R.~-M.~Ma, and T.~Li,
	``Optically Levitated Nanodumbbell Torsion Balance and GHz Nanomechanical Rotor,''
	Phys.\ Rev.\ Lett.\ {\bf 121}, 033603 (2018).
	
\bibitem{GHzRotor2}
  R.~Reimann, M.~Doderer, E.~Hebestreit, R.~Diehl, M.~Frimmer, D.~Windey, F.~Tebbenjohanns, and L.~Novotny,
	``GHz Rotation of an Optically Trapped Nanoparticle in Vacuum,''
	Phys.\ Rev.\ Lett.\ {\bf 121}, 033602 (2018).
	
\bibitem{Unruh2ndQuantization}
  W.~G.~Unruh,
	``Second quantization in the Kerr metric,''
	Phys.\ Rev.\ D {\bf 10}, 3194 (1974).		
	
\bibitem{Faces}
  J.~D.~Bekenstein and M.~Schiffer,
	``The many faces of superradiance,''
	Phys.\ Rev.\ D {\bf 58}, 064014 (1998).

\bibitem{GenSuperradiance}	
  M.~Richartz, S.~Weinfurtner, A.~J.~Penner, and W.~G.~Unruh,
 	``Generalised superradiant scattering,''
	Phys.\ Rev.\ D {\bf 80}, 124016 (2009).	
	
% Microwaves/Masers: wavelengths from 10cm (3 GHz) to 1mm (300 GHz), absorbed by polar molecules, coupling to vibrational and rotational modes, though often penetrates farther into surfaces than higher frequencies like infrared and light (this is why its used to cook food)
\bibitem{MilliLG1996}
  G.~A.~Turnbull, D.~A.~Robertson, G.~M.~Smith, L.~Allen, and M.~J.~Padgett,
	``The generation of free-space Laguerre-Gaussian modes at millimetre-wave frequencies by use of a spiral phaseplate,''
	Optics Comm.\ {\bf 127}, 127 (1996).
	
% Gyrotron (i.e. electron cyclotron maser) peaked 100 GHz, but can go 400 GHz and 1THz (for smaller wavelengths). In the millimetre-wave region, one has a tough time beam-splitting and beam-combining; diffraction is an issue for long-range propagation. High power output, though, for the reported gyrotron 	
\bibitem{GyrotronMasers2017}
  A.~Sawant, M.~S.~Choe, M.~Thumm, and E.~Choi,
	``Orbital Angular Momentum (OAM) of Rotating Modes Driven by Electrons in Electron Cyclotron Masers,''
	Sci.\ Rep.\ {\bf 7}, 3372 (2017).
	
% Radiowaves: wavelengths bigger than 10cm (3 GHz), typically 1m (300 MHz), 10m (30 MHz), 100m (3 MHz), 1km (300 kHz).

% Less than a GHz example. generated with vector antenna arrays. allow for the instantaneous local field vectors to be measured, and manipulated by software (not possible in regular optics)
\bibitem{LowFreqRadio2007}
  B.~Thid\'e, H.~Then, J.~Sj\"oholm, K.~Palmer, J.~Bergmann, T.~D.~Carozzi, Y.~N.~Istomin, N.~H.~Ibragimov, and R.~Khamitova,
	``Utilization of photon orbital angular momentum in the low-frequency radio domain,''
	Phys.\ Rev.\ Lett.\ {\bf 99}, 087701 (2007).

% General radio analysis for generation
\bibitem{OAMRadio}
  S.~M.~Mohammadi, L.~K.~S.~Daldorff, J.~E.~S.~Bergman, R.~L.~Karlsson, B.~Thid\'e, K.~Forozesh, T.~D.~Carozzi, and B.~Isham,
	``Orbital Angular Momentum in Radio - A System Study,''
	IEEE Trans.\ Antennas Propag.\ {\bf 58}, 565 (2010).
	
% Technically may be microwaves (around 94 GHz, or about 3mm), but called radio in paper. Supposed to be pretty cheap to make!	
\bibitem{GHzRadio}
  L.~Cheng, W.~Hong, and Z.~C.~Hao,
  ``Generation of Electromagnetic Waves with Arbitrary Orbital Angular Momentum Modes,''
	Sci.\ Rep.\ {\bf 4}, 4814 (2014).	
	
\bibitem{THz}
  J.~He, X.~Wang, D.~Hu, J.~Ye, S.~Feng, Q.~Kan, and Y.~Zhang,
		``Generation and evolution of the terahertz vortex beam,''
		Opt.\ Exp.\ {\bf 21}(17), 20230 (2013).
		
\bibitem{Doppler}	
  B.~A.~Garetz,
	``Angular Doppler effect,''
	J.\ Opt.\ Soc.\ Am.\ {\bf 71}(5), 609 (1981).
	
\bibitem{AngularAccel}
  Y.~Zhai, S.~Fu, C.~Yin, H.~Zhou, and C.~Gao,
	``Detection of angular acceleration based on optical rotational Doppler effect,''
	Opt.\ Exp.\ {\bf 27}(11), 15518 (2019).		
	
\bibitem{RotDopplerNonlinearOpt}
  G.~Li, T.~Zentgraf, and S.~Zhang,
	``Rotational Doppler effect in nonlinear optics,''
	Nat.\ Phys.\ {\bf 12}, 736 (2016). 
	
\bibitem{RotFreqShiftNew}
  M.~Michalski and W.~H\"uttner,
	``Experimental Demonstration of the Rotational Frequency Shift in a Molecular System,''
	Phys.\ Rev.\ Lett.\ {\bf 95}, 203005 (2005).
	
\bibitem{RotFreqShiftmm}
  J.~Courtial, D.~A.~Robertson, K.~Dholakia, L.~Allen, and M.~J.~Padgett,
	``Rotational Frequency Shift of a Light Beam,''
	Phys.\ Rev.\ Lett.\ {\bf 81}, 4828 (1998).
	
\bibitem{RotFreqShiftMeas}
  J.~Courtial, K.~Dholakia, D.~A.~Robertson, L.~Allen, and M.~J.~Padgett,
	``Measurement of the Rotational Frequency Shift Imparted to a Rotating Light Beam Possessing Orbital Angular Momentum,''
	Phys.\ Rev.\ Lett.\ {\bf 80}, 3217 (1998).	

	
\bibitem{Anton10000}
  R.~Fickler, G.~Campbell, B.~Buchler, P.~K.~Lam, and A.~Zeilinger,
	``Quantum entanglement of angular momentum states with quantum numbers up to 10010,''
	Proc.\ Natl.\ Acad.\ Sci.\ U.S.A. {\bf 113}(48), 13642 (2016).
	
\bibitem{Couple}
  H.~Shi and M.~Bhattacharya,
	``Coupling a small torsional oscillator to large optical angular momentum,''
	J.\ Mod.\ Opt.\ {\bf 00}(00), 1 (2012).	
	
%Anton
\bibitem{SmallBeamHighTopo}
  M.~Krenn, N.~Tischler, and A.~Zeilinger,
	``On small beams with large topological charge,''
	New J.\ Phys.\ {\bf 18}(3), 033012 (2016).	

\bibitem{SmallBeamHighTopo2}
  M.~Krenn and A.~Zeilinger,
	``On small beams with large topological charge: II. Photons, electrons and gravitational waves,''
	New J.\ Phys.\ {\bf 20}, 063006 (2018).	
		
% Most of the optomechanics references describe optical scalar vortex beams, for instance Laguerre-Gaussian modes. These nonetheless are topologically charged (recall that topological charge is the number of twists the fixed-phase wavefront of the vortex beam undergoes in one wavelength of propagation, and can be either positive or negative, depending on the handedness of the twist), so they have orbital angular momentum of m*hbar per photon.

% Alternatively, one can work with vectorial vortex beams, which are beams with polarization singularities. Included in this class of beams are the vector Bessel-Gauss modes, which possess cylindrical polarization symmetry.  

% "Perfect" vortices have intensity profiles that are independent of the topological charge (more accurately, the dark hollow radius does not depend on the topological charge). They are Fourier-transforms of the Bessel-Gauss modes (either the scalar Bessel-Gauss modes, which upon Fourier transforming yield perfect scalar vortex beams, or the vector Bessel-Gauss modes, which upon Fourier transforming yield perfect vector vortex beams).
\bibitem{PerfectHighTopo}
  Y.~Chen, Z.~X.~Fang, Y.~X.~Ren, L.~Gong, and R.~D.~Lu,
	``Generation and characterization of a perfect vortex beam with a large topological charge through a digital micromirror device,''
	Appl.\ Opt.\ {\bf 54}(27), 8030 (2015).

% Paper that introduced perfect vortices:
\bibitem{FirstPerfect}
  A.~S.~Ostrovsky, C.~Rickenstorff-Parrao, and V.~Arriz\'on,
	``Generation of the `perfect' optical vortex using a liquid-crystal spatial light modulator,''
	Opt.\ Lett.\ {\bf 38}, 534 (2013).
	
	% A better perfect vortex generation technique, by the people who introduced the idea of perfect vortices
\bibitem{SimplePerfect}
  J.~Garc\'ia-Garc\'ia, C.~Rickenstorff-Parrao, R.~Ramos-Garc\'ia, V.~Arriz\'on, and A.~S.~Ostrovsky,
	``Simple technique for generating the perfect optical vortex,''
	Opt.\ Lett.\ {\bf 39}, 5305 (2014).
	
\bibitem{PerfectHighOrder}
  N.~A.~Chaitanya, M.~V.~Jabir, and G.~K.~Samanta,
	``Efficient nonlinear generation of high power, high order, ultrafast `perfect' vortices in green,''
	Opt.\ Lett.\ {\bf 41}, 1348 (2016).

% Recent efficient method of generating perfect vortex beams
\bibitem{EffPerfVort2017}
  A.~A.~Kovalev, V.~V.~Kotlyar, and A.~P.~Porfirev,
	``A highly efficient element for generating elliptic perfect optical vortices,''
	Appl.\ Phys.\ Lett.\ {\bf 110}, 261102 (2017).

\bibitem{PerfectVectorScalarVortex}
  Y.~Liu, Y.~Ke, J.~Zhou, Y.~Lui, H.~Luo, S.~Wen, and D.~Fan,
	``Generation of perfect vortex and vector beams based on Pancharatnam-Berry phase elements,''
	Sci.\ Rep.\ {\bf 7}, 44096 (2017).
	
% Quantum possibilities: Optomechanics

\bibitem{RovibeEntangle2008}	
	M.~Bhattacharya, P.~L.~Giscard, and P.~Meystre,
	``Entangling the rovibrational modes of a macroscopic mirror using radiation pressure,''
	Phys.\ Rev.\ A {\bf 77}, 030303 (2008).

\bibitem{LaserRot2007}
  M.~Bhattacharya and P.~Meystre,
	``Using a Laguerre-Gaussian Beam to Trap and Cool the Rotational Motion of a Mirror,''
	Phys.\ Rev.\ Lett.\ {\bf 99}, 153603 (2007).
	
\bibitem{RotoMirrorEntangle2008}
  M.~Bhattacharya, P.~-L.~Giscard, and P.~Meystre,
	``Entanglement of a Laguerre-Gaussian cavity mode with a rotating mirror,''
	Phys.\ Rev.\ A {\bf 77}, 013827 (2008).
	
\bibitem{RotoStabilization}
  S.~J.~M.~Habraken and G.~Neinhuis,
	``Rotational stabilization and destabilization of an optical cavity,''
	Phys.\ Rev.\ A {\bf 79}, 011805(R) (2009).
	
\bibitem{LargeAngularMom}
H.~Shi and M.~Bhattacharya,
``Coupling a small torsional oscillator to large optical angular momentum,''
J.\ Mod.\ Optics {\bf 00}, 00, 1-8 (2012).	
	
\bibitem{THzRotor}
  A.~A.~Milner, A.~Korobenko, J.~W.~Hepburn, and V.~Milner,
	``Effects of Ultrafast Molecular Rotation on Collisional Decoherence,''
	Phys.\ Rev.\ Lett.\ {\bf 113}, 043005 (2014).
	
\bibitem{MacroRotSup}
  B.~A.~Stickler, B.~Papendell, S.~Kuhn, B.~Schrinski, J.~Millen, M.~Arndt, and K.~Hornberger,
	``Probing macroscopic quantum superpositions with nanorotors,''
	New J.\ Phys.\ {\bf 20}, 122001 (2018).	
	
\end{thebibliography}
\end{document}